\documentclass[5p]{elsarticle}

\usepackage{blindtext}
\usepackage{float}
\usepackage{amsmath}
\usepackage{amssymb}
\usepackage{graphicx}
\usepackage [english]{babel}
\usepackage{braket}
\usepackage[utf8]{inputenc}
\usepackage[pdfencoding=auto,psdextra]{hyperref}
\usepackage{url}

\usepackage{todonotes}

\begin{document}

\def\nuc#1#2{\relax\ifmmode{}^{#1}{\protect\text{#2}}\else${}^{#1}$#2\fi}
\title{Nuclear \emph{ab~initio} calculations of $^{6}\text{He}$ \(\beta\)-decay for beyond the Standard Model studies}

\author{Ayala Glick-Magid}
\address{Racah Institute of Physics, The Hebrew University, The Edmond J. Safra Campus - Givat Ram, Jerusalem 9190401, Israel}
\author{Christian Forss\'en}
\ead{christian.forssen@chalmers.se}
\address{Department of Physics, Chalmers University of Technology, SE-412 96 G\"oteborg, Sweden}
\author{Daniel Gazda}
\address{Nuclear Physics Institute, 25068 \v{R}e\v{z}, Czech Republic}
\author{Doron Gazit}
\ead{doron.gazit@mail.huji.ac.il}
\address{Racah Institute of Physics, The Hebrew University, The Edmond J. Safra Campus - Givat Ram, Jerusalem 9190401, Israel}
\author{Peter Gysbers}
\address{TRIUMF, 4004 Wesbrook Mall, Vancouver, British Columbia V6T 2A3, Canada}
\address{Department of Physics and Astronomy, University of British Columbia, Vancouver, British Columbia, V6T 1Z1, Canada}
\author{Petr Navr\'atil}
\address{TRIUMF, 4004 Wesbrook Mall, Vancouver, British Columbia V6T 2A3, Canada}

\begin{abstract}
  Precision measurements of $\beta$-decay observables offer the possibility to search for deviations from the Standard Model. A possible discovery of such deviations requires accompanying first-principles calculations. Here we compute the nuclear structure corrections for the $\beta$-decay of $^6$He which is of central interest in several experimental efforts. We employ the impulse approximation together with wave functions calculated using the \emph{ab~initio} no-core shell model  with potentials based on chiral effective field theory. We use these state-of-the-art calculations to give a novel and comprehensive analysis of theoretical uncertainties. We find that nuclear corrections, which we compute within the sensitivity of future experiments, create significant deviation from the naive Gamow-Teller predictions, making their accurate assessment essential in searches for physics beyond the Standard Model.
\end{abstract}

\maketitle

The Standard Model (SM) is very successful in explaining the vast majority of observed phenomena in particle physics.
Nevertheless, several important questions remain unanswered and the search for evidence of Beyond the Standard Model (BSM) physics is one of the heralds of contemporary particle physics~\cite{peskin2016trail}.
In particular, recent years have brought advances in the sensitivity of \(\beta\)-decay studies, and several high precision experimental efforts~\cite{GONZALEZALONSO2019165,cirigliano2019precision} have been deployed, as a ``precision frontier'' to search for BSM physics---alternative to the high-energy frontier represented by the Large Hadron Collider (LHC)~\cite{CIRIGLIANO201393}. \(\beta\)-decay observables are sensitive to interference of currents of SM particles and hypothetical BSM physics. Such couplings are proportional to $\big(v/\Lambda\big)^2$, with $v\approx 174$~GeV, the SM vacuum expectation value, and $\Lambda$ the new physics energy scale. This entails that a difference between SM theoretical predictions and experiment that can be inferred as a result of a $ \sim 10^{-4}$ coupling between SM and BSM physics would suggest new physics at a scale that is out of the reach of current particle accelerators.

However, discovering such minute deviations from the SM predictions demands also high-precision theoretical calculations. In the case of $\beta$-decays this challenge entails nuclear structure calculations with quantified uncertainties.
The field of \emph{ab~initio} nuclear structure calculations has significantly evolved in the last two decades. Concomitantly, model-independent nuclear interactions based on chiral effective field theory ($\chi$EFT)~\cite{epelbaum2009,machleidt2011} have been developed. A showcase example of this progress, relevant for studies of weak interactions with nuclei, is the recent accurate calculations of $\beta$-decay rates for nuclei with masses up to $A=100$~\cite{Gysbers2019NatPhys}.

In this Letter we present a theoretical analysis of $\beta$-decay observables of the pure Gamow--Teller (GT) transition  $^{6}\text{He}\left(0^{+}_\text{gs}\right){\to}{}^{6}\text{Li}\left(1^{+}_\text{gs}\right)$, with an endpoint energy of $Q=3.50521(5)$~MeV\footnote{The $n$ digits within the parenthesis specify the theoretical precision as $\pm$ these digits added to the last $n$ digits of the value presented.}~\cite{osti_1774904}.

Our motivation to study $^6$He is twofold: First, $^6$He is a light nucleus for which \emph{ab~initio} many-body calculations are numerically tractable, and thus can test feasibility to reach the precision needed by experiments to constrain BSM signatures. Second, $^6$He is being studied in several ongoing, or soon to be initiated, experimental campaigns at Laboratoire de Physique Corpusculaire de CAEN (France)~\cite{cirigliano2019precision}, National Superconducting Cyclotron Laboratory (USA) ~\cite{doi:10.1063/1.4955362}, the University of Washington CENPA (USA) by the He6-CRES Collaboration~\cite{PhysRevLett.114.162501}, and at SARAF accelerator (Israel)~\cite{Ohayon2018}. The SARAF experiment is focusing on the angular correlation between the emitted $\beta$-particles, while the other aforementioned experiments will measure the \(\beta\)-electron energy spectrum.
As the experiments aim for a per-mil level of accuracy, precise calculations based on the SM are needed.

The $\beta$-decay spectrum of a pure GT transition takes the simple form: $d\omega \propto1+a_{\beta\nu}\vec{\beta}\cdot\hat{\nu}+b_{\text{F}}\frac{m_{e}}{E}$~\cite{PhysRev.106.517}. Here, $\vec{\beta}=\frac{\vec{k}}{E}$, while $m_{e}$, $E$ and $\vec{k}$ are the mass, energy and momentum of the emitted $\beta$-electron, respectively, and $\vec{\nu} = \nu \hat{\nu}$ is the momentum of the emitted anti-neutrino. $a_{\beta\nu}$ is the angular correlation coefficient between the emitted electron and anti-neutrino, and $b_{\text{F}}$ is the so-called Fierz interference term.
The $V-A$ structure of the weak interaction within the SM entails that $a_{\beta\nu}=-\frac{1}{3}$ and $b_{\text{F}}=0$ for pure GT transitions~\cite{GONZALEZALONSO2019165}, neglecting other effects. These values can be modified in the presence of BSM physics, but also by nuclear-structure corrections. Thus, disregarding nuclear structure effects might lead to wrong interpretation of the experiments.
Notice that $b_{\text{F}}$ is linear in BSM couplings such that constraining $b_{\text{F}} < 10^{-3}$ in a GT transition yields a sensitivity to BSM currents characterized by tensor coupling $\epsilon_T < 1.5\cdot 10^{-4}$, or new physics at a scale $\Lambda > 14$~TeV~\cite{cirigliano2019precision,CIRIGLIANO201393,Falkowski2021} . On the other hand, $a_{\beta\nu}$ is proportional to $\epsilon_T^2$~\cite{lee1956question,PhysRev.106.517}, thus demanding higher precision to reach the same BSM constraints.
The extraction of both these parameters from $^6$He $\beta$-decay demands two experimental settings as $b_F$ is obtained from spectral measurements in which the angular correlation term vanishes. This is in contrast to the case of first-forbidden unique transitions, where a simultaneous extraction of $a_{\beta\nu}$ and $b_{\text{F}}$ is possible from the \(\beta\)-energy spectrum~\cite{GLICKMAGID2017285}.

Currently, the state-of-the-art measurement of $^6$He angular correlations is a recoil ion energy measurement from 1963~\cite{PhysRev.132.1149}, resulting in $a_{\beta\nu}$ value consistent with $-\frac{1}{3}$ up to the experimental error of 0.9\%. Corrections to this result were offered over the years by adding radiative corrections~\cite{GLUCK1998493}, and updating the $^{6}\text{He}$ shake-off probability~\cite{PhysRevA.92.050701} and Q-value~\cite{PhysRevLett.108.052504, PhysRevA.82.042513}. The present work is, to our knowledge, the first consistent calculation of nuclear-structure related corrections to these observables, taking into account the full nuclear dynamics, shown as important already in Ref.~\cite{CALAPRICE1976301}, and using $\chi$EFT to quantify systematic uncertainties in the nuclear modeling. Nuclear-structure effects are particularly important since they entail a finite $b_{\mathrm{F}}^{1^+ \beta^-}$ value which in turn has been shown to distort the extraction of $a_{\beta\nu}$~\cite{PhysRevC.94.035503}. In the following, we use \emph{ab~initio} calculations of nuclear wave functions and the weak transition matrix elements to predict experimentally relevant observables. We focus on corrections related to nuclear structure, and use 
a recent $\beta$-decay formalism~\cite{glickmagid2021formalism}, to augment the point prediction with a theoretical uncertainty estimate.

The full expression for \nuc{6}{He} \(\beta^-\)-decay differential distribution within the SM---including the leading shape and recoil corrections, i.e., next to leading order (NLO) in GT---takes the following form:
\begin{equation}
\begin{split}
  \label{eq:spect}
  \frac{d\omega^{1^+ \beta^-}}{dE\frac{d\Omega_{k}}{4\pi}\frac{d\Omega_{\nu}}{4\pi}}=\frac{4}{\pi^{2}}\left(E_0-E\right)^{2}kE F^{-}\!\left(Z_{f},E\right) C_{\text{corr}} \left\lvert\braket{\Vert \hat{L}_{1}^{A}\Vert} \right\rvert^{2} \\
  \times 3 \left(1+\delta_{1}^{1^{+}\beta^{-}}\right)
  \left[1+a_{\beta\nu}^{1^{+}\beta^{-}}\vec{\beta}\cdot\hat{\nu} +b_{\text{F}}^{1^{+}\beta^{-}}\frac{m_{e}}{E}\right] ,
\end{split}
\end{equation}
where $E_0$ is the maximal electron energy, $k=\lvert\vec{k}\rvert$, $\braket{\Vert \hat{L}^A_{1}\Vert}$ is the reduced matrix element between the initial-{} and final-state wave functions of the rank-$1$ spherical tensor longitudinal operator of the axial current (proportional to the GT operator). In general, we use the superscript $A(V)$ to denote axial-(polar-)vector contribution to the weak nuclear current. Furthermore, $F^{-}\!\left(Z_{f},E\right)$ is the Fermi function (calculated here according to~\cite{Venkataramaiah1985}), which takes into account the deformation of the \(\beta\)-particle wave function due to the long-range electromagnetic interaction with the nucleus, and $C_{\text{corr}}$ represents other corrections which do not originate purely in the weak matrix element, such as radiative corrections, finite-mass and electrostatic finite-size effects, and atomic effects. These corrections are assumed to be well known~\cite{RevModPhys.90.015008} and are not taken into account in the following calculations as they do not affect the observables.

Taking recoil and shape corrections into account, the $\beta-\nu$ correlation becomes
\begin{equation}
  \label{eq:angcorr}
  a_{\beta\nu}^{1^+ \beta^-} =-\frac{1}{3}\left(1+\tilde{\delta}_{a}^{1^{+}\beta^{-}}\right)\text{.}
\end{equation}
Similarly, an $m_e/E$ spectral behavior appears, similar to the BSM-induced Fierz interference term, via a nuclear-structure dependent factor
\begin{equation}
  \label{eq:bfierz}
  b_{\mathrm{F}}^{1^+ \beta^-}
  =\delta_{b}^{1^{+}\beta^{-}}.
\end{equation}

These relative corrections originate from rank-1 multipole operators, $\hat{C}_{1}^{A}$ (axial charge) and $\hat{M}_{1}^{V}$ (vector magnetic or weak magnetism), and can be written as
\begin{equation}
  \label{eq:corrections}
  \begin{split}
    \delta_{1}^{1^{+}\beta^{-}} &\equiv\frac{2}{3}\mathfrak{Re}\left[
      -E_0
      \frac{\braket{\Vert \hat{C}_{1}^{A}/q\Vert} }{\braket{\Vert\hat{L}_{1}^{A}\Vert} }
      +\sqrt{2}\left(E_0-2E\right)
      \frac{\braket{\Vert \hat{M}_{1}^{V}/q\Vert} }{\braket{\Vert \hat{L}_{1}^{A}\Vert} }
    \right]\\
    &- \frac{4}{7} E R \alpha Z_f
    -\frac{233}{630}\left(\alpha Z_f\right)^2 
    , \\
    \tilde{\delta}_{a}^{1^{+}\beta^{-}} &\equiv\frac{4}{3}\mathfrak{Re}\left[ 2E_0\frac{\braket{\Vert \hat{C}_{1}^{A}/q\Vert} }{ \braket{\Vert \hat{L}_{1}^{A}\Vert} }+\sqrt{2}\left(E_0-2E\right)
      \frac{\braket{\Vert \hat{M}_{1}^{V}/q\Vert} }{ \braket{\Vert \hat{L}_{1}^{A}\Vert} }
    \right]\\
    &+ \frac{4}{7} E R \alpha Z_f - \frac{2}{5}E_0 R\alpha Z_f 
    , \\
    \delta_{b}^{1^{+}\beta^{-}} &\equiv\frac{2}{3}m_{e}\mathfrak{Re}\left[ \frac{\braket{ \Vert \hat{C}_{1}^{A}/q\Vert} }{\braket{ \Vert \hat{L}_{1}^{A}\Vert} }
      +\sqrt{2}
      \frac{\braket{ \Vert \hat{M}_{1}^{V}/q\Vert} }{\braket{ \Vert \hat{L}_{1}^{A}\Vert} } \right],
  \end{split}
\end{equation}
where $\vec{q}=\vec{k}+\vec{\nu}$ is the momentum transfer, $R$ is the radius of the nucleus, $\alpha \approx \frac{1}{137}$ is the fine structure constant, and $Z_f=3$ is the charge of the final nucleus. In the nomenclature of~\cite{glickmagid2021formalism}, $\hat{C}_{1}^{A}$ and $\hat{M}_{1}^{V}$ are dominated by two small dimensionless parameters $\epsilon_{\text{NR}}\epsilon_{qr}$ and $\epsilon_{\text{recoil}}$. In the current kinematics, the non-relativistic small parameter $\epsilon_{\text{NR}} \sim P_{\text{Fermi}}/m_{N}\approx0.2$ ($P_{\text{Fermi}}$ is Fermi momentum and $m_{N}$ is the nucleon mass), $\epsilon_{qr}\sim qR \approx 0.05$ and $\epsilon_{\text{recoil}}\sim q/m_{N}\approx 0.004$ while subleading (NNLO in GT) corrections to Eq.~\eqref{eq:corrections} would be of the order of $\frac{1}{15}\epsilon_{qr}^{2} \sim 10^{-4}$~\footnote{This arises from the residual correction of the electric multipole operator $\hat{E}^A_{J}$ from its low energy approximation, i.e., the residual is the second term in the exact relation $\hat{E}^A_{J M_J} = \sqrt{\frac{J+1}{J}} \hat{L}^A_{J M_J} -i\sqrt{\frac{2J+1}{J}}\int d^3r j_{J+1}\left(qr\right) \vec{Y}_{J J+1 1}^{M_J}\left(\hat{r}\right)\cdot \vec{J^A}\left(\vec{r}\right)$, where $\vec{J^A}$ is the hadron axial current, $\vec{Y}_{Jl1}^{M_J}$ is the vector spherical harmonic, and the spherical Bessel function $j_{J}\left(\rho\right) \approx \frac{\rho^J}{\left(2J+1\right)!!}$~\cite{glickmagid2021formalism}.}.

Non-radiative electromagnetic corrections, arising from the distortion of the electron wave function, can be divided into three main contributions~\cite{glickmagid2021formalism}. First, spectrum corrections, which are taken into account for the leading order by including the Fermi function \(F^{-}\!\left(Z_{f},E\right)\), and for the sub-leading orders 
by the terms proportional to the order of $\epsilon_{qr} \epsilon_{c} \sim 10^{-3}$ (with $\epsilon_{c} \sim \alpha Z_f \approx 2\cdot 10^{-2}$) and $\epsilon_{c}^2\sim 5\cdot 10^{-4}$~\cite{PhysRevC.19.1467},
incorporated in Eq.~\eqref{eq:corrections}~\cite{RevModPhys.90.015008, hayen2020consistent}. Our numerical calculations in combination with the expressions in Refs.~\cite{RevModPhys.90.015008, hayen2020consistent} yield that sub-sub-leading orders for the spectrum corrections of \nuc{6}{He} are $\lesssim 10^{-4}$, and this is factored into our error estimation.
Second, correction terms to the multipole operators are proportional to $\epsilon_{\text{recoil}} \epsilon_{qr} \epsilon_{c} \sim 5\cdot10^{-6}$ ~\cite{PhysRevC.19.1467}, and therefore smaller than other corrections and than the sensitivity of current experiments, and thus are not included in the calculations.
Third, the electrostatic gauge field results in a correction $\Delta E_c$ to the energy transfer $E_0$, i.e., the difference between the Coulomb potentials of the decaying and final nucleus~\cite{behrens1982electron}, and is discussed following Eq.~\eqref{eq:mapping}.

The main nuclear structure dynamics are encapsulated both in the
nuclear wave functions and in the structure of the multipole
operators, which are expansions of the hadronic currents and charges
within the nucleus.
At the low energies characterizing nuclear $\beta$-decays, this
dynamics, microscopically governed by quantum chromodynamics (QCD), can be effectively reduced into a field theory of nucleons, pions and short-range interactions by the use of $\chi$EFT.
This results in a consistent expansion governed by a small parameter $\epsilon_{\rm EFT}$, which dictates the accuracy of the theory.  Below we estimate that $\epsilon_{\rm EFT} \lesssim 0.15$ for the present study.
A detailed derivation of the power-counting of electro-weak operators in $\chi$EFT can be found in Ref.~\cite{Krebs_EPJA_2020} and references therein. 
Weak probes generally interact with currents of ever growing clusters of nucleons. However, within $\chi$EFT, interactions with currents of bigger clusters are suppressed. For weak magnetism, $\hat{M}_{1}^{V}$, the two-body current part is suppressed by $\epsilon_{\rm EFT}$ compared to the leading-order single-nucleon current, while the $\hat{L}_{1}^{A}$ and $\hat{C}_{1}^{A}$ two-body current terms are associated with the next order, $\epsilon_{\rm EFT}^2$.

We calculate the needed multipole operators within the so-called impulse approximation, i.e., single-nucleon currents weakly interacting with the \(\beta\) particles, while neglecting two- and higher-body currents. 
In this approximation, the three nuclear operators \(\hat{L}^A\), \(\hat{C}^A\) and \(\hat{M}^V\) appearing in Eqs.~\eqref{eq:spect} and \eqref{eq:corrections} can be expressed in terms of four basic multipole operators \(\hat{\Sigma}^{\prime\prime}\), \(\hat{\Omega}^{\prime}\), \(\hat{\Delta}\), and \(\hat{\Sigma}^\prime\)~\cite{Donnelly:1979ezn}
(see ~\ref{app:multipoles} for definitions) as
\begin{equation}
  \label{eq:mapping}
  \begin{split}
    \frac{\hat{C}_{J M_J}^{A}}{q} &= \sum_{j=1}^{A} \frac{i}{m_N}
    \left[g_A \hat{\Omega}^{\prime}_{J M_J} (q \vec{r}_j)\right.\\
    &\left. - \frac{1}{2}\frac{\tilde{g}_P}{2m_N}
      \left(E_0 + \Delta E_c\right)  \hat{\Sigma}_{J M_J}^{\prime\prime}(q \vec{r}_j)
    \right] \tau^{+}_j,\\
    \hat{L}_{J M_J}^{A} &= \sum_{j=1}^{A} i \left(g_A+\frac{\tilde{g}_P}{\left(2m_N\right)^2}
    q^2\right)
    \hat{\Sigma}_{J M_J}^{\prime\prime} (q \vec{r}_j)
    \,\tau^{+}_j, \\
    \frac{\hat{M}_{J M_J}^{V}}{q} &= \sum_{j=1}^{A} \frac{-i}{m_N}
    \left[ g_V \hat{\Delta}_{J M_J} (q \vec{r}_j)-\frac{1}{2}
      \mu \hat{\Sigma}^{\prime}_{J M_J} (q \vec{r}_j) \right]
    \tau^{+}_j.
  \end{split}
\end{equation}
Here, $J$ $(M_J)$ is the multipole rank (projection), $m_N$ is the nucleon mass,
\(\vec{r}_j\) \((\tau^{+}_j)\) is the \(j\)th nucleon position vector (isospin-raising operator). Note that the sum runs over the \(A\) nucleons (not to be confused with the $A$ labeling axial quantities). 
In~\eqref{eq:mapping}, $\mu \approx 4.706$ is the nucleon isovector magnetic moment; $g_V = 1$, $g_A= -1.2756\left(13\right)$~\cite{10.1093/ptep/ptaa104} and $\tilde{g}_P = -\frac{\left(2 m_N\right)^2}{m_{\pi}^2 - q^2} g_A$~\cite{WALECKA1975113} are the hadronic vector, axial-vector and pseudo-scalar charges, which correspond to the nucleon form factors at $q=0$.
In general, the nucleon form factors include momentum-transfer corrections proportional to $q_\mu^2$~\cite{GONZALEZALONSO2019165}. These, however, are $\sim \frac{1}{6}\epsilon_{qr}^2$, and thus smaller than the needed precision.

As aforementioned, the correction $\Delta E_c$ to the energy transfer $E_0$ is the difference between the Coulomb energies of the final and initial states of the nucleus, $\Delta E_c \equiv \langle {}^{6}\text{Li}\text{ }1^{+}_\text{gs} \lvert V_c \rvert {}^{6}\text{Li}\text{ }1^{+}_\text{gs} \rangle - \langle {}^{6}\text{He}\text{ }0^{+}_\text{gs} \lvert V_c \rvert {}^{6}\text{He}\text{ }0^{+}_\text{gs} \rangle$, where $V_c$ denotes the full Coulomb potential operator.
Quantum Monte Carlo calculations in Ref.~\cite{PhysRevC.56.1720}
present the Coulomb energy difference $\Delta E_c=0.85\left(3\right)$~MeV, which is the value that we use in our calculations.
This result is consistent with the experimental value for the Coulomb displacement energy between a pair of isobaric analog levels, which is given by
$\Delta E_c \equiv M_{Z>}-M_{Z<}+\Delta_{nH} = M\left({}^{6}\text{Li}\text{ }0^{+}\right)-M\left({}^{6}\text{He}\text{ }0^{+}_\text{gs}\right)+\Delta_{nH} = 0.837(10)$~MeV~\cite{ANTONY19971}, where $M_{Z>}$ ($M_{Z<}$) is the atomic mass of the higher (lower) Z member of the analog pair, and $\Delta_{nH}$ is the neutron–hydrogen mass difference.

Wave functions of \nuc{6}{He} and \nuc{6}{Li} and the many-body matrix elements of the multipole operators in~\eqref{eq:corrections} are obtained within the \emph{ab~initio} no-core shell model (NCSM)~\cite{PhysRevLett.84.5728,PhysRevC.62.054311,BARRETT2013131} using $\chi$EFT interactions as the only input.
We utilized two chiral interactions in this work, namely NNLO${}_{\mathrm{opt}}$~\cite{N2LOopt} and NNLO${}_{\mathrm{sat}}$~\cite{N2LOsat}.
The former was constructed from $\chi$EFT at the NNLO order with inclusion of only the \(N\!N\) terms. This interaction reproduces reasonably well the experimental binding energies ($\sim 5\%$) and radii for $A=3, 4$ nuclei, as well as for the $A=6$ systems that are relevant for this work~\cite{Forssen:2017wei}. The NNLO${}_{\mathrm{sat}}$ interaction is also constructed at the NNLO order of $\chi$EFT---but includes \(3N\) forces, and is more accurate for heavier systems~\cite{Calci16,Ka16,Ha16,Arthuis2020}.

The present calculations are performed using a Slater determinant \(A\)-nucleon harmonic-oscillator (HO) basis in the $M$-scheme. The basis is characterized by the HO frequency $\Omega$ and contains up to $N_{\rm max}$ HO excitations above the lowest Pauli-principle-allowed configuration. We apply the standard procedure of introducing one-body transition densities to compute matrix elements of one-body operators between initial-{} and final-state NCSM wave functions as
\begin{equation}
  \label{eq:obtdobme}
  \begin{split}
    \langle \Psi_f \Vert \sum_{j=1}^{A} \hat{O}_{J}(\vec{r}_j) \Vert \Psi_i
    \rangle & = \frac{-1}{\sqrt{2 J + 1}} \sum_{\lvert\alpha\rvert,
      \lvert\beta\rvert} \langle \lvert\alpha\rvert \Vert
    \hat{O}_J(\vec{r}) \Vert \lvert\beta\rvert \rangle\\
    &\times \langle \Psi_f
    \Vert
    (a^\dagger_{\lvert\alpha\rvert}\tilde{a}_{\lvert\beta\rvert})_J
    \Vert \Psi_i \rangle.
  \end{split}
\end{equation}
The operator matrix elements \(\langle \lvert\alpha\rvert \Vert \hat{O}_J(\vec{r}) \Vert \lvert\beta\rvert\rangle\), reduced in the angular momentum, are evaluated between HO states which depend on the coordinate \(\vec{r}\) and are labeled by their nonmagnetic quantum numbers \(\lvert\alpha \rvert\ (\lvert\beta\rvert)\). In Eq.~\eqref{eq:obtdobme}, \(\tilde{a}_{\lvert\beta\rvert, m_j} = (-1)^{j_\beta - m_\beta} a_{\lvert\beta\rvert, -m_j}\), with \(a^\dagger_{\alpha}\) and \(a_\beta\) the creation and annihilation operators for the single-particle HO states \(\ket{\alpha}\) and \(\ket{\beta}\), respectively, coupled to the angular momentum \(J\). In the present case we have $\ket{\Psi_i}=\ket{{\rm ^6He}\, 0 _{\rm gs}^+ 1}$, $\ket{\Psi_f}=\ket{{\rm ^6Li}\, 1 _{\rm gs}^+ 0}$, and $J=1$.

However, the single-particle coordinates \(\vec{r}_j\) and \(\vec{r}\) in Eq.~\eqref{eq:obtdobme} are measured with respect to the center of the HO potential. Thus, these matrix elements clearly contain contributions from spurious center-of-mass (CM) motion. Fortunately, the exact factorization of NCSM wave functions into a product of the physical intrinsic eigenstate and a CM state in the $0\hbar\Omega$ excitation makes it possible to remove the effect of the CM completely. Therefore, we introduce a translationally-invariant one-body density depending on coordinates and momenta measured from the CM of the nucleus, e.g., $\vec{\xi}=-\sqrt{A/(A-1)}(\vec{r}-\vec{R}_{\rm CM})$. This density is obtained as a direct generalization of the radial translationally-invariant density~\cite{Navratil2004} by considering dependence on nucleon spins. Specifically, we replace the standard density~\eqref{eq:obtdobme} by
\begin{multline}
  \label{eq:trinvobd}
  \langle \Psi_f \Vert \sum_{j=1}^{A} \hat{O}_{J}(\vec{r}_j-\vec{R}_{\mathrm{CM}}) \Vert \Psi_i \rangle
  = \frac{-1}{\sqrt{2 J + 1}} \\
  \times \sum_{\lvert a\rvert \lvert
    b\rvert\lvert\alpha\rvert \lvert\beta\rvert} \langle \lvert
  a\rvert \Vert
  \hat{O}_J(-\sqrt{A-1/A}\vec{\xi}) \Vert \lvert b\rvert \rangle \\
  \times (M^J)^{-1}_{\lvert a\rvert \lvert
    b\rvert,\lvert\alpha\rvert \lvert\beta\rvert} \langle \Psi_f \Vert
  (a^\dagger_{\lvert\alpha\rvert}\tilde{a}_{\lvert\beta\rvert})_J
  \Vert \Psi_i \rangle ,
\end{multline}
with the $M^J$ matrix and further details given in Ref.~\cite{PhysRevC.104.064322}. The ``one-body'' HO states $\ket{a(b)}$ depend on the Jacobi coordinate $\vec{\xi}$ as opposed to the single-particle HO states $\ket{\alpha(\beta)}$ that depend on single-particle coordinates $\vec{r}$.

\begin{figure}[tb]
  \centering
  \includegraphics[width = \columnwidth]{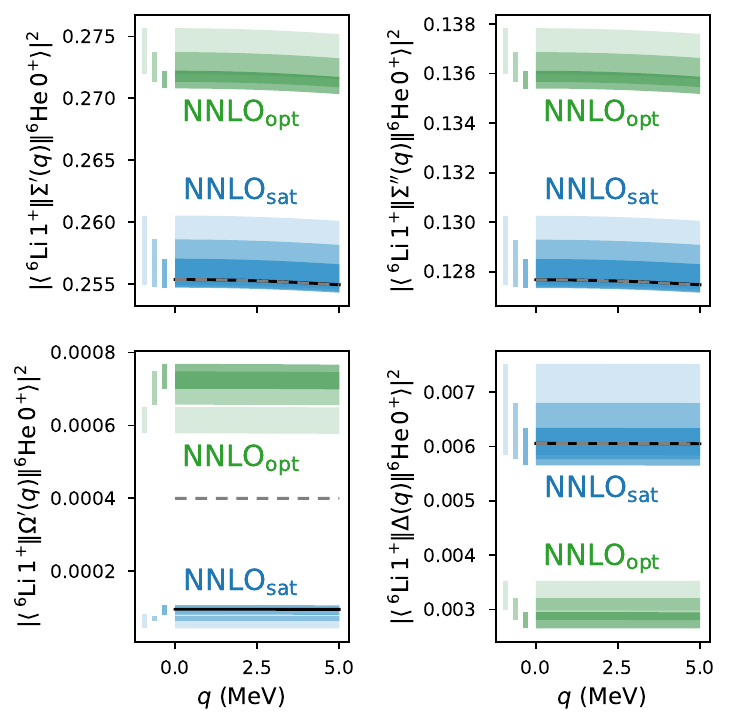}
  \caption{Dependence of nuclear matrix elements on the NCSM model space parameters and nuclear Hamiltonians. Light, medium, and heavy filled bands correspond to $N_\mathrm{max}=8, 10, 12$ for the NNLO$_\mathrm{sat}$ interaction including 3NF (blue bands), and $N_\mathrm{max}=10, 12, 14$ for the NNLO$_\mathrm{opt}$ interaction with only 2NF (green bands). The width of the bands show the variation with HO frequency $\hbar\Omega = 16, 20, 24$~MeV. The solid (dashed) line shows the result with the NNLO$_\mathrm{sat}$ interaction at $N_\mathrm{max} = 12$, $\hbar\Omega = 20$~MeV computed with translationally-invariant (standard) one-body densities.\label{fig:nmelements}}
\end{figure}
Results for the nuclear matrix elements of the one-body basic multipole operators \(\hat{\Sigma}^{\prime\prime}\), \(\hat{\Omega}^{\prime}\), \(\hat{\Delta}\), and \(\hat{\Sigma}^\prime\) are shown in Fig.~\ref{fig:nmelements}. These matrix elements are then used to construct the nuclear structure input, $\hat{L}^A_1, \hat{C}^A_1, \hat{M}^V_1$, as in Eq.~\eqref{eq:mapping}. The convergence in terms of the basis frequency $\hbar\Omega$ and the model space truncation $N_\mathrm{max}$ is well controlled, but it is clear that results depend somewhat on the nuclear Hamiltonian. We also note that the translationally-invariant one-body density, Eq.~\eqref{eq:trinvobd}, and the standard one-body density, Eq.~\eqref{eq:obtdobme}, give the same many-body matrix elements at $q=0$ for the $\hat{\Sigma}^\prime$, $\hat{\Sigma}^{\prime\prime}$, and $\hat{\Delta}$ operators while the many-body matrix elements of $\hat{\Omega}^\prime$ differ. In particular, the spurious center-of-mass component of the wave functions contaminates the matrix elements when the gradient in the first term of $\hat{\Omega}^\prime$ is applied on the wave function. With an increasing $q$, all the operators become contaminated by spurious center-of-mass contributions although the effect is quite small for $\hat{\Sigma}^\prime$, $\hat{\Sigma}^{\prime\prime}$, and $\hat{\Delta}$, i.e., it is not visible on the resolution scale of Fig.~\ref{fig:nmelements}.

\begin{figure}[h]
  \centering
  \includegraphics[width = \columnwidth]{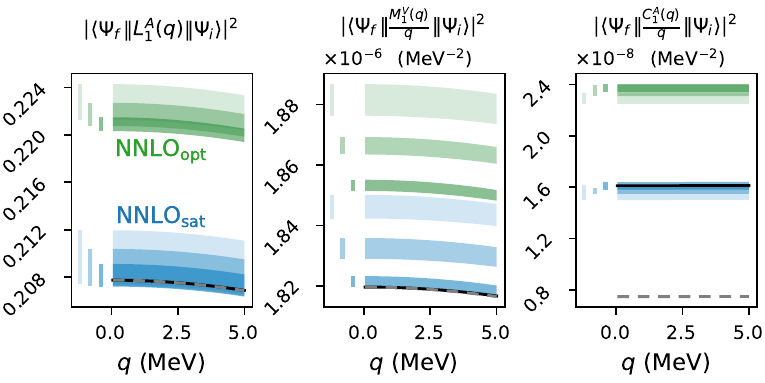}
  \caption{Dependence of nuclear matrix elements on the NCSM model-space parameters and nuclear Hamiltonians. The light to dark bands correspond to $N_\mathrm{max}=8, 10, 12$ for the NNLO${}_{\mathrm{sat}}$ (blue) interaction and $N_\mathrm{max}=10, 12, 14$ for the NNLO${}_{\mathrm{opt}}$ (green) interaction. The width of the bands show the variation with HO frequency $\hbar\Omega = 16, 20, 24$~MeV. The solid (dashed) line shows the result with the NNLO${}_{\mathrm{sat}}$ interaction at $N_{\mathrm{max}} = 12$, $\hbar\Omega = 20$~MeV computed with translationally-invariant (standard) one-body densities.\label{fig:CLM}}
\end{figure}

\begin{figure}[htb]
  \centering
  \includegraphics[width = \columnwidth]{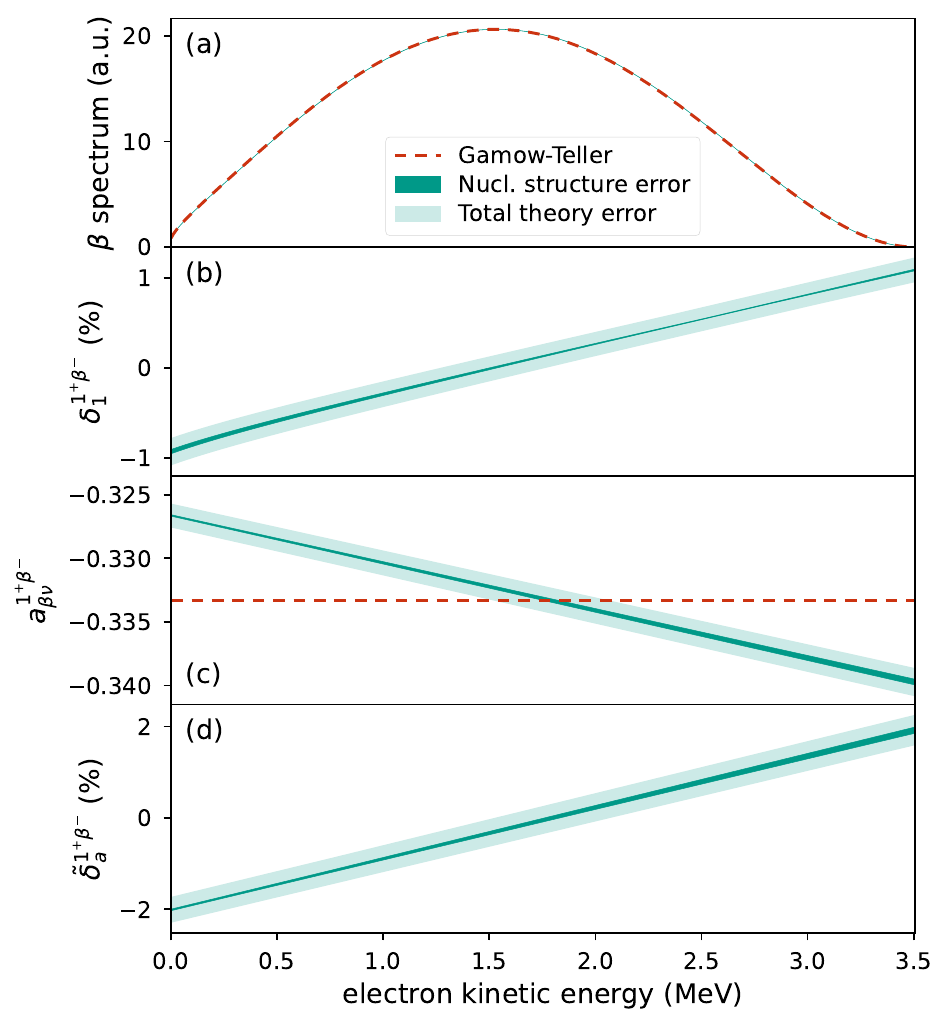}
  \caption{(a) Calculated energy dependence of the spectrum of \nuc{6}{He} \(\beta\)-decay, in arbitrary units. Dashed line is the pure GT spectrum, while the filled bands include nuclear-structure dependent corrections. (b) The residual nuclear structure correction $\delta_{1}^{1^{+}\beta^{-}}$ compared to the pure GT spectrum (Eq.~\eqref{eq:corrections}).
    (c) Energy dependence of the angular correlation $a_{\beta\nu}$
    from Eq.~\eqref{eq:angcorr}. Dashed line corresponds to the SM value, $a_{\beta\nu}^{\rm GT}=-1/3$. (d) Relative size of the
    $\tilde{\delta}_{a}^{1^{+}\beta^{-}}$ correction from~\eqref{eq:corrections}.
    The width of the dark filled bands shows the variation with the employed nuclear Hamiltonian and NCSM model space parameters for HO frequency $\hbar\Omega = 16, 20, 24$~MeV, $N_\mathrm{max}=8, 10, 12$ ($10, 12, 14$) using the NNLO${}_{\mathrm{sat}}$ (NNLO${}{_\mathrm{opt}}$) interaction, using translationally-invariant one-body densities. The width of the light filled band shows the total estimated theory error.\label{fig:spect}}
\end{figure}

The $\nuc{6}{He}{\to}\nuc{6}{Li}$ nuclear matrix elements that appear in Eq.~\eqref{eq:corrections} are shown in Fig.~\ref{fig:CLM}. We note that results are indeed sensitive to the removal of spurious CM components as performed in this work.
The technology for evaluating these $q$-dependent multipole matrix elements with NCSM wave functions was developed in Ref.~\cite{Gazda:2016mrp} based on the work in~\cite{Fitzpatrick:2012ix}. We study the convergence as a function of model space parameters and the dependence on the nuclear Hamiltonian as shown in Fig.~\ref{fig:CLM}. These nuclear-structure uncertainties are propagated to the final BSM-related observables considered in this work, see the dark filled bands in Fig.~\ref{fig:spect}, and shown to be small. 
Overall, the most sophisticated description is achieved by the $N_{\rm max}{=}12$ NNLO$_{\rm sat}$ calculation with the correction of the CM effect and the HO frequency of 20~MeV. At that frequency, the $^6$He and $^6$Li g.s.\ energies are at their minimum.

As aforementioned, the lack of two-body currents in the multipole operators leads to an absence in the theory, dominated by a small parameter $\epsilon_{\rm EFT}$, where the $\hat{M}_{1}^{V}$ two-body current part is characterized by $\epsilon_{\rm EFT}$, while $\hat{L}_{1}^{A}$ and $\hat{C}_{1}^{A}$ two-body current terms are proportional to $\epsilon_{\rm EFT}^2$. To verify this, and estimate these EFT uncertainties better, we make use of other observables, where higher-order EFT calculations were compared to experiment, namely the \nuc{6}{Li} magnetic moment and M1 transition, and the \nuc{6}{He} half-life. According to Ref.~\cite{PhysRevC.87.035503} (\cite{PhysRevLett.126.102501}), the two-body current, which is an NLO correction for this operator, has a vanishing contribution to the \nuc{6}{Li} magnetic moment, and a 20\% (10\%) contribution to the $\nuc{6}{Li} ( 0^{+}{\to} 1^{+})$ B(M1) transition. As B(M1) contains the squared matrix element, this entails at most a 10\% two-body-current contribution for $\hat{M}_{1}^{V}$. Additionally, we compared our $\hat{L}_{1}^{A}$ calculations with the empirical GT operator, $\lvert\rm GT\left(^{6} \rm He\right)\rvert_{\rm expt}=2.161(5)$, calculated from the \nuc{6}{He} half-life~\cite{PhysRevC.79.065501}, and found a deviation of $1.5\%$ in $\hat{L}_{1}^{A}$, consistent with the fact that these corrections are of higher order in EFT counting than $\hat{M}_{1}^{V}$. This consistency allows us to conservatively estimate that $\epsilon_{\rm EFT} \lesssim 0.15$.

As shown in Fig.~\ref{fig:spect}b(d), we find up to 1\% (2\%) corrections to the \(\beta\) spectrum (angular correlation), consistent with the a~priori estimates based on the small parameters of the problem. However, these corrections depend on the electron kinetic energy, thus extracting $a_{\beta\nu}$ requires an energy-weighted average, adhering to the particular experimental setup. Here, we exemplify the important effect of this procedure, by using an average of $a_{\beta\nu}$ weighted by the spectrum $\frac{d\omega^{1^{+}\beta^{-}}}{dE}$. In this example, the total correction to $a_{\beta\nu}$ due to nuclear structure is
\begin{equation}
  \label{eq:delta_a_result}
  \left<\tilde{\delta}_{a}^{1^{+}\beta^{-}}\right> = -2.54\left(68\right)\cdot 10^{-3},
\end{equation}
i.e., a 7 per-mil correction to the SM $a_{\beta \nu}^{\text{GT}}=-\frac{1}{3}$.
This, however, is a naive value, as one should keep in mind the (often neglected) dependence of the measured $a_{\beta\nu}$ value on the $b_{\text{F}}$-analogous term detailed below.

Such a term with a similar spectral behavior as the Fierz interference can be extracted from the corrected spectrum, and our calculations indicate that it is non-zero
\begin{equation}
  \label{eq:b_result}
  b_{\mathrm{F}}^{1^+ \beta^-}
  =\delta_{b}^{1^{+}\beta^{-}} = -1.52\left(18\right)\cdot10^{-3}.
\end{equation}
This result, with an uncertainty of $\sim 10^{-4}$, is vital for ongoing experiments, aiming for a per-mil level of precision.

In order to extract the $\beta-\nu$ correlation coefficient $a_{\beta\nu}$, one notices that the spectral shape suggests that $a_{\beta\nu}^{\text{measured}}=\frac{a_{\beta\nu}} {1+b_{\text{F}}\left<\frac{m_{e}}{E}\right>}$~\cite{PhysRevC.94.035503}, resulting in the following relation:
\begin{equation}
\label{eq:a_measured}
  \begin{split}
    a_{\beta\nu} &=
    a_{\beta\nu}^{\text{measured}}
    -a_{\beta \nu}^{\text{GT}}\left(
    \left<\tilde{\delta}_{a}^{1^{+}\beta^{-}}\right>
    - b_{\text{F}}^{1^{+}\beta^{-}} \left<\frac{m_{e}}{E}\right> \right) \\
    &= a_{\beta\nu}^{\text{measured}} - 0.70\left(24 \right) \cdot 10^{-3},
    \end{split}
  \end{equation}
where $\left<\frac{m_{e}}{E}\right>= 0.28536\left(10\right)$.

However, a realistic measurement cannot probe directly the correlation between the neutrino and the $\beta$ particle. For example, in the 1963 experiment~\cite{PhysRev.132.1149}, the recoil ion energy spectrum was studied, resulting in a different effect. The effect  for \nuc{6}{He} is given by $a_{\beta\nu}^{\text{measured}}= a_{\beta\nu}+0.127\, b_{\text{F}}$~\cite{PhysRevC.94.035503}, so the measured value (including radiative corrections~\cite{GLUCK1998493}, influence of the updated shake-off probability~\cite{PhysRevA.92.050701} and Q-value~\cite{PhysRevLett.108.052504, PhysRevA.82.042513}) $a_{\beta\nu}^{\text{measured}} + \delta_{\text{rad,so,Q}}=-0.3324(30)$ should be modified to
\begin{equation}
\label{eq:a_Carlson}
  \begin{split}
    a_{\beta\nu} & =
    a_{\beta\nu}^{\text{measured}} + \delta_{\text{rad,so,Q}}
    -\left(a_{\beta \nu}^{\text{GT}}
    \left<\tilde{\delta}_{a}^{1^{+}\beta^{-}}\right>
    +0.127 b_{\text{F}}^{1^{+}\beta^{-}}\right) \\
    & = -0.3331\left(32\right)
    \text{.}
    \end{split}
\end{equation}

Thus, the extracted $a_{\beta\nu}$ depends on corrections that imitate the spectral dependence of the Fierz term (suppressed by a numerical factor of about $0.1$). Importantly, this indirectly induces a linear dependence of this observable upon BSM corrections, beyond the naive quadratic dependence of $a_{\beta \nu}$. Consequently, $\sim 10^{-4}$ experimental precision on this observable would entail tighter BSM constraints~\cite{mishnayot2021constraining}.

Summarizing, we have used a $\chi$EFT framework combined with the \emph{ab initio} NCSM to analyze the nuclear-structure related corrections to \nuc{6}{He} \(\beta\)-decay observables. In particular, we have studied the angular correlation coefficient and a nuclear structure term with an inverse energy spectral dependence, imitating a Fierz interference term. Our analysis uses the existence of small parameters, originating mainly in the low-energy regime characterizing $\beta$-decays, to quantify the relevant theoretical uncertainties. We find that the induced $m_e/E$ behavior, that can be wrongly interpreted as a result of Fierz interference between SM and BSM currents, is significantly different than the naive SM value of zero. Our theoretical prediction comes with less than 15\% uncertainty.
Furthermore, 0.2 per-mil bounds were found for SM nuclear structure effects correcting the angular correlation coefficient. Albeit these are smaller than the current experimental uncertainty, future angular correlations measurements of \nuc{6}{He} decay, aimed at reducing the current error by one order of magnitude, should use these bounds to check for BSM signatures, due to the indirect dependence of the angular correlations on the Fierz term.
These results increase significantly the potential to correctly check the SM, as well as pin-pointing possible deviations from it.

\section*{Acknowledgments}
This work was initiated as a result of the stimulating environment at the ECT* workshop ``Precise beta decay calculations for searches for new physics'' in Trento. We wish to acknowledge the support of the ISF grant no.\ 1446/16 (DGazit and AGM), the Swedish Research Council, Grant Nos.\ 2017-04234 (CF and DGazda) and 2021-04507 (CF), the Czech Science Foundation GA\v{C}R grants Nos.\ 19-19640S and 22-14497S (DGazda), and the NSERC Grants No.\ SAPIN-2016-00033 (PG and PN) and PGSD3-535536-2019 (PG). AGM’s research was partially supported by a scholarship sponsored by the Ministry of Science \& Technology, Israel. TRIUMF receives federal funding via a contribution agreement with the National Research Council of Canada.  Computing support came from an INCITE Award on the Summit supercomputer of the Oak Ridge Leadership Computing Facility (OLCF) at ORNL, and from Westgrid and Compute Canada. Parts of the computations and data handling were enabled by resources provided by the Swedish National Infrastructure for Computing (SNIC) at Chalmers Centre for Computational Science and Engineering (C3SE), the National Supercomputer Centre (NSC) partially funded by the Swedish Research Council through grant agreement no.~2018-05973. Additional computational resources were supplied by the project ``e-Infrastruktura CZ'' (e-INFRA CZ LM2018140) supported by the Ministry of Education, Youth and Sports of the Czech Republic and IT4Innovations at Czech National Supercomputing Center under project number~OPEN-24-21~1892.

\appendix

\section*{Appendices}

\section{Nuclear multipole operators%
\label{app:multipoles}}

The four basic operators from SM electroweak theory that appear in Eq.~\eqref{eq:mapping} in the main text are defined as~\cite{Donnelly:1979ezn}
\begin{equation}
  \label{eq:operators}
  \begin{split}
    \hat{\Sigma}^{\prime\prime}_{JM_J}(q \vec{r}_j) &= \left[
      \frac{1}{q} \vec{\nabla}_{\vec{r}_j} M_{JM_J}(q
      \vec{r}_j)  \right] \cdot \vec{\sigma}_j,\\
    \hat{\Omega}^\prime_{JM_J}(q \vec{r}_j) &= M_{JM_J}(q \vec{r}_j)
    \,\vec{\sigma}_j \cdot \vec{\nabla}_{\vec{r}_j} +
    \frac{1}{2}\hat{\Sigma}^{\prime\prime}_{JM_J}(q \vec{r}_j),\\
    \hat{\Delta}_{JM_J}(q \vec{r}_j) &= \vec{M}_{JJM_J}(q \vec{r}_j)
    \cdot \frac{1}{q}\vec{\nabla}_{\vec{r}_j}, \\
    \hat{\Sigma}^{\prime}_{JM_J}(q \vec{r}_j) &= -i \left[ \frac{1}{q}
      \vec{\nabla}_{\vec{r}_j} \times \vec{M}_{JJM_J}(q
      \vec{r}_j) \right] \cdot \vec{\sigma}_j,
  \end{split}
\end{equation}
with \(\vec{\sigma}_j\) being the Pauli spin matrices associated with nucleon \(j\). Furthermore, \(M_{JM_J}(q \vec{r}_j) = j_J(q r_j) Y_{JM_J}(\hat{r}_j)\) and \(\vec{M}_{JLM_J}(q \vec{r}_j) = j_L(q r_j) \vec{Y}_{JLM_J}(\hat{r}_j)\),
where $j_{J}$ are the spherical Bessel functions, $Y_{J M_J}$ ($\vec{Y}_{Jl1}^{M_J}$) are the spherical harmonics (vector spherical harmonics), and $J$ $(M_J)$ is the multipole rank (projection).

When evaluating the one-body-like Jacobi-coordinate matrix elements appearing in Eq.~\eqref{eq:trinvobd} in the main text we first carry out the gradients in the parenthesis of $\hat{\Sigma}^\prime$, $\hat{\Sigma}^{\prime\prime}$ (see, e.g., Refs.~\cite{Donnelly:1979ezn,HAXTON2008345}) and then replace $\vec{r}$ by $-\textstyle{\sqrt{\frac{A-1}{A}}}\vec{\xi }$ in all the operators, and, in addition, we replace the gradients (momenta) in $\hat{\Omega}^\prime$ and $\hat{\Delta}$ by $ -\textstyle{\sqrt{\frac{A-1}{A}}}\vec{\nabla}_{\vec{\xi }}$.

We note that one-body matrix elements of the seven basic multipole operators for electroweak processes can be carried out analytically in the HO basis as demonstrated in Refs.~\cite{Donnelly:1979ezn,HAXTON2008345}. In Ref.~\cite{HAXTON2008345}, a Mathematica script is provided for the calculation of the matrix elements. These results can be readily applied to calculate the matrix elements of the translationally-invariant versions of the operators we use here. In the analytic results, e.g., Eqs.~(17)--(19) in Ref.~\cite{HAXTON2008345}, (i) the $q$ is replaced by $-\textstyle{\sqrt{\frac{A-1}{A}}} q$, (ii) the matrix elements (18) and (19) in Ref.~\cite{HAXTON2008345} are multiplied by one more factor of $-\textstyle{\sqrt{\frac{A-1}{A}}}$ due to the gradient (momentum) acting on the wave function, and, finally, (iii) yet another factor of $-\textstyle{\sqrt{\frac{A-1}{A}}}$ is applied to terms with $1/q$, i.e., $\hat{\Sigma}^\prime$, $\hat{\Sigma}^{\prime\prime}$, and $\hat{\Delta}$ (\ref{eq:operators}), to compensate for the extra scaling in step (i).

\section{Comparison to literature%
\label{app:Calaprice}}

We would like to compare our results to the ones presented in the original 1975 Calaprice calculation~\cite{PhysRevC.12.2016}, which is following the notation of Holstein and Treiman. According to that notation, there are three nuclear form factors needed to describe the beta-dacay transition to first order in recoil: the Gamow-Teller $c$, the weak magnetism $b$, and the induced tensor $d$.
These can be connected, at leading order, to the matrix elements we calculated in Eq.~\eqref{eq:obtdobme} (in the main text), through the following leading-order relations ~\cite{behrens1982electron}:
\begin{equation}
  \label{eq:Calprice}
  \begin{split}
  c_1 &\cong 2 \sqrt{3 \pi} g_A \braket{\Vert\sum_{j=1}^A \tau^+_j \hat{\Sigma}^{\prime\prime}_1 \Vert}, \\
  b &\cong -2 \sqrt{6 \pi} A \braket{\Vert \sum_{j=1}^A \tau^+_j \left( g_V \hat{\Delta}_1 - \frac{1}{2} g_M\hat{\Sigma}^{\prime}_1 \right) \Vert}, \\
  d^{I} &\cong 2 \sqrt{3 \pi} A g_A \braket{\Vert\sum_{j=1}^A \tau^+_j  \left(\hat{\Omega}^{\prime}_1 -\frac{1}{2}\hat{\Sigma}^{\prime\prime}_1\right)\Vert}.
\end{split}
\end{equation}
We found that $c_1 \cong 2.85\left(14\right)$, in agreement with $c \cong 2.75$ of Calaprice.
Also the magnetic form factor $b \cong 66.7\left(8.3\right)$ we obtained is in agreement with $b=69.0\left(1.0\right)$ experimental value that Calaprice presents. As we mentioned, the uncertainty of the magnetic multipole is large because it is dominated by $\epsilon_{\text{EFT}}$. However, in $\left<\tilde{\delta}_{a}^{1^{+}\beta^{-}}\right>$ the contribution of the magnetic multipole is averaged out.
Last, unlike the Calaprice values, our calculations result in a negative value for $\frac{d^{I}}{Ac_1}$. Using the leading-order relation from Eq.~\eqref{eq:Calprice}, we obtain $\frac{d^{I}}{Ac_1} \cong -0.45\left(4\right)$. If we use the exact same operator as Calaprice, then we get $\frac{d^{I}}{Ac_1} \cong -0.29$. This value itself is somehow between the theoretical value 0.12, and the experimental value $2.0\left(1.5\right)$ presented by Calaprice.

\bibliography{main}

\end{document}